\documentclass[a4paper]{jpconf}
\usepackage{graphicx}
\usepackage{amsmath, amsfonts, amsthm, amssymb, graphicx}
\usepackage{subfig}
\usepackage{epstopdf}

\usepackage{fontenc}
\usepackage{textcomp}

\def\be{\begin{equation}}
\def\ee{\end{equation}}
\def\ba{\begin{eqnarray}}
\def\ea{\end{eqnarray}}

\def\bdm{\begin{displaymath}}
\def\edm{\end{displaymath}}

\def\bq{\begin{quote}}
\def\eq{\end{quote}}

\def\ltap{\ \raise.3ex\hbox{$<$\kern-.75em\lower1ex\hbox{$\sim$}}\ }
\def\gtap{\ \raise.3ex\hbox{$>$\kern-.75em\lower1ex\hbox{$\sim$}}\ }
\def\gl{\ \raise.5ex\hbox{$>$}\kern-.8em\lower.5ex\hbox{$<$}\ }
\def\roughly#1{\raise.3ex\hbox{$#1$\kern-.75em\lower1ex\hbox{$\sim$}}}

 at 10truept

\newcommand{\beq}{\begin{equation}}
\newcommand{\eeq}{\end{equation}}
\newcommand{\bea}{\begin{eqnarray}}
\newcommand{\eea}{\end{eqnarray}}
\newcommand{\beqa}{\begin{eqnarray}}
\newcommand{\eeqa}{\end{eqnarray}}

\begin{document}
\title{The good, the bad and the ugly .... of Ho\v rava gravity}

\author{Antonio Padilla}

\address{School of Physics \& Astronomy, University of Nottingham, Nottingham NG7 2RD, UK}

\ead{antonio.padilla@nottingham.ac.uk}

\begin{abstract}
I review the good, the bad and the ugly of the non-projectable versions of Ho\v rava gravity. I explain how this non-relativistic  theory was constructed and  why it was touted with such excitement as a quantum theory of gravity. I then review some of the issues facing the theory, explaining  how strong coupling occurs and why this is such a problem for both phenomenology and the question of renormalisability. Finally I  comment on possible violations of Equivalence Principle, and explain why these could be an issue for Blas {\it et al}'s "healthy extension".  This paper was presented as a talk at PASCOS 2010 in Valencia.

\end{abstract}

\section{Introduction}
Let us begin with a few words regarding the title. A couple of days after I gave this talk in Valencia, Blas {\it et al} submitted the following paper to the arXiv: {\it Models of non-relativistic quantum gravity: the good, the bad and the healthy} \cite{gbh}. At the time I did not know whether the similarity with the title of my talk was deliberate, or mere coincidence. My initial suspicion was that it had to be deliberate since  what they call healthy, I have called ugly! However, I have since been assured by the authors of \cite{gbh} that it was nothing more than a rather amusing coincidence. 

By now Ho\v rava gravity comes in many different forms \cite{Horava1, Blas2, Horava2}. From now on, when I refer to ``Ho\v rava gravity" I will mean the so-called {\it non-projectable} theories for which the lapse function can depend on space $N=N(x, t)$. I will begin by explaining what is so good about Ho\v rava's original idea, and why his toy model was  touted with such excitement as a renormalisable theory of quantum gravity. The key idea is to break the isotropic scaling of  space and time, allowing us  to improve the UV behaviour of the theory without introducing ghosts. To do this we must break diffeomorphism invariance, and this is where the trouble starts.  By breaking diffeomorphism invariance we introduce an additional scalar degree of freedom. What happens to this scalar mode as we approach the classical GR  limit in the infra-red? Does it decouple or does it become strongly coupled? Unfortunately it is the latter \cite{me1, Blas1}. This is the bad of Ho\v rava gravity. Filling a conspicuous gap in  the existing literature, we will explain precisely why this strong coupling is so bad for both the renormalisability of the theory and its phenomenology.

Following these developments,  Blas {\it et al} proposed an extension of Ho\v rava's theory which they claimed did not run into problems with strong coupling, and should be a phenomenologically viable toy model of quantum gravity  \cite{Blas2, Blas3}. This is what they refer to as the ``healthy" extension, and I have referred to as ``ugly".  Perhaps ``ugly" is a bit harsh. What I find unappealing is that they introduce a  large dimensionless parameter ($\sim 10^7$) by hand, through a very low scale of Lorentz violation. This large parameter is absolutely crucial in order to solve the problems  described in the previous paragraph\cite{Blas2, Blas3}. As we will explain, the situation may even be worse than this since the low scale of Lorentz violation might lead to unacceptably large violations of the Equivalence Principle \cite{me2}. Establishing whether or not this is the case is a work in progress \cite{me3}. Should we find that Blas {\it et al}'s model \cite{Blas2} can solve the strong coupling problem without any additional phenomenological problems, then I think the large parameter is a small price to pay for a phenomenologically consistent  toy model of quantum gravity. Beauty, as they say, is in the eye of the beholder! For now, we should be patient and wait for the results of \cite{me3}.

\section{The Good}
To understand what is so good about Ho\v rava's original idea, we need to understand the trouble with gravity. Classical gravity is well described by General Relativity. At the quantum level, GR is non-renormalisable, essentially because  the coupling constant has negative mass dimension, $[G_N]=-2$, and the graviton propagator scales like $1/k^2$. To see ways in which we might get round this, let us consider the following toy model, 
\be
{\cal L}=-\frac{1}{2} (\partial \phi)^2+\lambda \phi^6
\ee
As with GR, the propagator scales like  $1/k^2$, and the coupling constant has mass dimension $[\lambda]=-2$, so the theory looks like it might well be non-renormalisable. One way to fix this might be to add relativistic higher derivative corrections to improve the UV behaviour of the propagator. Schematically, 
\be
\frac{1}{k^2}\to \frac{1}{k^2}+ \frac{1}{k^2} \lambda k^4  \frac{1}{k^2}+  \frac{1}{k^2} \lambda k^4  \frac{1}{k^2} \lambda k^4  \frac{1}{k^2}+\ldots=\frac{1}{k^2-\lambda k^4}
\ee
After canonically normalising in the UV, $\phi \to \hat \phi/\sqrt{|\lambda|}$, we get a new coupling constant $\hat \lambda=\lambda^{-2}$ whose mass dimension is non-negative. The theory appears to be power counting renormalisable, but it has come at a price -- the higher derivatives introduce additional ghost-like degrees of freedom. This  is easily  seen by rewriting the propagator as 
\be
\frac{1}{k^2-\lambda k^4}=\frac{1}{k^2}-\frac{1}{k^2-\lambda^{-1}}
\ee
and is patently unacceptable. Note that the problem really lies with higher order time derivatives as opposed to space derivatives. This observation suggests a possible resolution.  Let us abandon Lorentz invariance and introduce higher order spatial derivatives without introducing any higher order time derivatives.The former should improve the UV behaviour of the propagator, whereas the latter guarantees the absence of ghosts. To this end, we  modify the kinetic term
\be
-\frac{1}{2} (\partial \phi)^2 \to \frac{1}{2} \dot \phi^2-\frac{1}{2} \phi (-\Delta)^z \phi, \label{zphi}
\ee
where $\Delta$ is the spatial Laplacian. We now have non-relativistic dispersion relation $w^2 \propto p^{2z}$, or in other words, time and space scale differently, $[x]=-1, ~[t]=-z$. For large enough $z$, it follows that the coupling constant has a non-negative scaling dimension, $[\lambda]=4z-6$. As an example, consider  the  $z=3$ theory: we   expect it to be power counting renormalisable, but we do not expect ghosts. Of course, the price we have paid is to have broken Lorentz invariance, which is well tested at low energies.  However, we can cope with this by adding a relevant operator of the form ${\cal L}_{rel}=\frac12 c^2 \phi \Delta \phi$. This leaves the good UV physics unaffected, but allows Lorentz invariance to be restored as an emergent symmetry in the IR, with an emergent speed of light $c$.

Ho\v rava's nice idea is to apply these methods to gravity \cite{Horava1}. First we must abandon Lorentz invariance, which means choosing a preferred time, $t$ and making an ADM split
\be
ds^2=-N(x, t)^2c^2 dt^2+g_{ij}(x, t)(dx^i+N^i(x, t) dt)(dx^j+N^j(x, t) dt)
\ee
As usual,  $g_{ij}$ is the spatial metric, $N^i$ is the shift vector, and $N$ is the lapse function. By allowing $N$ to depend on space we note that we are dealing with the {\it non-projectable} version of the theory. Now we see that we no longer have full diffeomorphism invariance. It has been replaced by ``foliation preserving diffeomorphisms"
\be
x^i \to \tilde x^i=\tilde x^i(x, t), \qquad t \to \tilde t=\tilde t(t)
\ee
The action is constructed from the following covariant objects: the spatial metric $g_{ij}$ and the extrinsic curvature $K_{ij}=\frac{1}{2N}(\dot g_{ij}-2\nabla_{(i} N_{j)})$, where $\nabla_i$ is the spatial covariant derivative.  In principle, one could also include terms with $\nabla_i \log N$ \cite{Blas2}, but let us come back to that in section \ref{sec:ug}. To build the gravitational analogue of the action (\ref{zphi}) we first replace the kinetic term
\be
\frac12 \dot \phi^2 \to \frac{1}{\kappa} \sqrt{g}N (K_{ij}K^{ij}-\lambda K^2)
\ee
where $\kappa$ is the gravitational coupling with scaling dimension $[\kappa]=z-3$, and $\lambda$ is a dimensionless parameter that also runs with scale. Clearly, for the $z=3$ theory the gravitational coupling constant is dimensionless and we might expect the theory to be power counting renormalisable!  Sticking with $z=3$, we build the remaining part of the UV action as follows:
 \be \label{pot}
 -\frac12 \phi (-\Delta)^3 \phi \to   \beta \kappa \sqrt{g}N \nabla_k R_{ij} \nabla^k R^{ij} + \ldots
 \ee
where $\beta$ is a dimensionless parameter and  $R_{ij}$ is the spatial Ricci tensor. Here ``$\ldots$" denotes any  of the other possible dimension $6$ operators that    one might wish to include in the potential. Just as we needed to  restore Lorentz invariance for the low energy scalar, now we need to be able to restore GR at low energies in the gravitational theory. To do this we add a relevant operator
\be
{\cal L}_{rel}=\frac{c^2}{\kappa}\sqrt{g}N R
\ee
The full $z=3$ gravitational theory is now given by
\be
S=\int dt d^3 x   ~\frac{1}{\kappa} \sqrt{g}N (K_{ij}K^{ij}-\lambda K^2+c^2 R)+\text{higher order spatial derivatives}
\ee
If we neglect the higher order spatial derivatives (as we would in the IR), we are left with something that looks very similar to the action for GR,
\be S_{GR}=\frac{c^4}{ 16 \pi G_N } \int dt d^3 x   ~ \sqrt{g}N \left(\frac{K_{ij}K^{ij}-K^2}{c^2}+R\right)
\ee
The claim then is that $\lambda$ flows to $1$ at low energies, such that we recover GR with an emergent speed of light, $c$, and an emergent Newton's constant, $G_N =\kappa c^2/16\pi$. It would seem that we have a theory of gravity that is well behaved at high energies, and recovers GR at low energies!

\section{The Bad} \label{sec:bad}
As the cynic might say, if something is too good to be true, it is probably a lie. For Ho\v rava gravity, it turns out that there are some  concerns of varying degrees of seriousness. At the not-so-serious end we have to contend with the enormous number of possible terms to be included in the potential -- in other words, the ``$\ldots$" present in equation (\ref{pot}). Ho\v rava tried to reduce the number by borrowing the principle of {\it detailed balance} from Condensed Matter Theory \cite{Horava1}, although this has since been ruled out phenomenologically \cite{me1}.  Whilst this is merely an aesthetic consideration, there is a more serious issue involving light cones and fine tuning. We have argued that Lorentz invariance is not exact, and emerges in the IR due to the presence of relevant operators. However, there is no reason to expect different particle species to see the same emergent light-cones. For this to happen we need to fine-tune the coefficients of the relevant operators in each case. A third concern relates to the formal structure of the theory as opposed to phenomenology -- the constraint algebra is dynamically inconsistent. This manifests itself through the lapse vanishing asymptotically for  generic solutions to the constraint equations\cite{hen}. 

The most serious issue of all stems from the breaking of diffeomorphism invariance.  To understand why, we briefly return to our scalar example. Here we broke Lorentz invariance, which is just a symmetry of the {\it background} and does not affect the number of degrees of freedom.  In contrast, for our gravity theory we broke diffeomorphism invariance, which is the  {\it dynamical} symmetry of General Relativity.  We therefore expect new dynamical degrees of freedom to appear. Now, the critical question is: what happens to these additional degrees of freedom as we approach the so-called ``GR limit" at low energies? I can think of just two possibilities: either they decouple (which is good), or they become strongly coupled (which is bad).

In Ho\v rava gravity it turns out to be the latter \cite{me1, Blas1}. Fluctuations about Minkowski space suggest that there exists an additional scalar mode that becomes strongly coupled at an energy scale of around $\Lambda_{naive} \sim \sqrt{\frac{c^3 |1-\lambda|}{\kappa}} \sim M_{pl}c^2 \sqrt{|1-\lambda|}$ \cite{me1}.  {\bf This means that the extra mode is strongly coupled on all scales in the ``GR limit" where $\lambda \to 1$}.  In fact, the situation is even worse than this.  A more careful analysis \cite{Blas1} including fluctuations on general curved backgrounds yields a strong coupling scale of the order $\Lambda_{strong} \sim \Lambda_{naive}^\frac{3}{4} (c/L)^\frac{1}{4} $, where $L$ measures the characteristic length scale of the background\footnote{The background spatial curvature $\bar R_{ij} \sim 1/L^2$ and the background extrinsic curvature $\bar K_{ij} \sim c/L$.} . This means that strong coupling kicks in at all scales not just in the ``GR limit" ($\lambda \to 1$), but also in the Minkowski limit ($L \to \infty$).

In each case, the dynamics of the extra mode was revealed using the Stuckelberg trick \cite{stuck}. Let us remind ourselves of the spirit of this trick by means of a simpler example. The massless photon is described by a Lagrangian ${\cal L}= -\frac{1}{4} F_{\mu\nu}^2$ and  has just two degrees of freedom due to gauge invariance, $A_\mu \to A_\mu+\partial_\mu \psi$. Now consider a theory of  massive photons (with some interactions)
\be
{\cal L}=-\frac{1}{4} F_{\mu\nu}^2-\frac{m^2}{2}A_\mu^2 +A_\mu J^\mu
\ee
 As we no longer have gauge invariance, we allow $\partial_\mu J^\mu \neq 0$ and note that the massive photon picks up an extra degree of freedom. What happens to this extra mode as $m \to 0$? To reveal this we perform a Stuckelberg trick, artificially restoring gauge invariance by means of the field redefinition $A_\mu=\tilde A_\mu+\frac{1}{m}\partial_\mu \phi$. The new action is given by
 \be
 {\cal L}=-\frac{1}{4} \tilde F_{\mu\nu}^2-\frac12 (\partial \phi)^2 +J^\mu \left(\tilde A_\mu +\frac{1}{m}\partial_\mu \phi\right) +{\cal O}(m)
 \ee
 and is manifestly invariant under $\tilde A_\mu \to \tilde A_\mu+\partial_\mu \psi, ~\phi \to \phi-m\psi$. As $m \to 0$, the current-scalar interaction diverges, or in other words, the Stuckelberg field, $\phi$, becomes strongly coupled in the massless limit!
 
 The Stuckelberg trick in Ho\v rava gravity works along the same lines, by artificially restoring missing gauge invariance \cite{me1, Blas1}. Indeed, we restore diffeomorphism invariance by redefining the ADM slicing in terms of a Stuckelberg field, $\phi(x, t)$. That is the slices go from $t=$ constant $\to$ $\phi(x, t)=$ constant. It is this Stuckelberg field that becomes strongly coupled at the scale $\Lambda_{strong}$.
 
What is so bad about strong coupling? Is it bad for renormalisability? Not necessarily. It is well known that QED becomes strongly coupled in the UV due to the presence of a Landau pole, and yet the theory is still renormalisable. However,  Ho\v rava gravity is not so well behaved. To understand this, consider why we believed the theory to be renormalisable in the first place.   We inferred a schematic form of the action in terms of perturbative degree of freedom, simply by identifying curvatures with derivatives of the graviton,   e.g. $R_{ij} \leadsto \nabla^2 h_{ij}$. The problem is that this completely ignores the Stuckelberg field. Furthermore, the Stuckelberg action looks nothing like the renormalisable actions studied in \cite{Visser1}.  Crucially, the dispersion relation for the Stuckelberg field does {\it not} scale like $z=3$ in the UV \cite{Blas3, me2} so we no longer expect our theory to be renormalisable.

Is strong coupling bad for phenomenology? Again, not necessarily. In DGP gravity, for example, strong coupling is linked to the so-called Vainshtein effect which helps to screen an extra scalar on solar system scales \cite{dgp}. The problem with Ho\v rava gravity is that strong coupling occurs over nearly all scales close to the ``GR limit".  Strong coupling can be associated with a breakdown of perturbation theory, and so perturbation theory will hardly ever apply in Ho\v rava gravity. What this means is that the effective perturbative degrees of freedom of GR are no longer applicable --the true degrees of freedom will be bound states of the graviton and Stuckelberg fields. Without knowing the precise form of those bounds states, we lose some predictive power, but is the theory ruled out? I suspect it is. The reason is that perturbative GR is actually very well tested by the binary pulsar observations of Hulse and Taylor \cite{pulsars}. It is difficult to see how these can be reproduced given that a perturbative description hardly ever applies.

I conclude this section with the following orthogonal observation: Ho\v rava gravity may suffer from violations of the Equivalence Principle \cite{me2}.  There are strict bounds \cite{eotvos} on the size of these violations which one should be able to use to rule out regions of parameter space. To understand how the violations may arise, we need to understand how to couple matter in these theories. In the relativistic Stuckelberg picture, the matter action $S_m[\Psi_n; \gamma_{\mu\nu}, \phi]$ depends on the matter fields, $\Psi_n$, the spacetime metric, $\gamma_{\mu\nu}$, and  the Stuckelberg field, $\phi$. In the absence of full diffeomorphism invariance, we no longer have the usual conservation of energy-momentum.   Instead, using the foliation preserving diffeomorphisms, we can show that \cite{me2}
\be
h_{\alpha \nu} \nabla_\mu T^{\mu\nu}=0, \qquad \frac{1}{\sqrt{-\gamma}}\frac{\delta S_m}{\delta \phi}=-\frac{n_\nu \nabla_\mu T^{\mu\nu}}{\sqrt{-(\nabla \phi)^2}}
\ee
where $n_\mu=\partial_\mu \phi/\sqrt{-(\nabla \phi)^2}$ is the unit normal to the spatial slices, and $h_{\mu\nu}=\gamma_{\mu\nu}+n_\mu n_\nu$ is the spatial projector. Thus, non-conserved sources can carry a form of Stuckelberg ``charge", $\Gamma \sim \nabla T^{\mu\nu}/T^{\mu\nu}$.

 With this in mind, we considered the field due to a slowly varying point mass with $T^{\mu\nu}=M\exp(-t\Gamma_\text{source} ) \delta^3(\vec x)\text{diag}(1, 0, 0,0)$ \cite{me2}. This has Stuckelberg charge $\Gamma_\text{source}$. Probe particles also carrying Stuckelberg charge feel an extra Stuckelberg force, that dominates beyond a critical radius, $r_c \sim (\lambda-1)/\sqrt{ \Gamma_\text{probe} \Gamma_\text{source}}$. In this Stuckelberg region, different probes with different $
\Gamma$s will fall at different rates, in violation of the Equivalence Principle. Such violations are measured by the Eotvos parameter \cite{eotvos}, $\eta \sim \frac{\Gamma_1-\Gamma_2}{\Gamma_1+\Gamma_2}$, which  can be large even for small probe charges $\Gamma_1$ and $\Gamma_2$. To avoid these problems, we need to push the critical radius out to large values so that the Stuckelberg force is subdominant on astrophysical scales. This means taking  large $\lambda$ or small $\Gamma$s. With large $\lambda$ we deviate too much from GR and face   problems with Lorentz violation \cite{gbh}.  We therefore require small Stuckelberg charges. In fact, one might be tempted to assume that all sources are conserved so that all Stuckelberg charges vanish. Whilst this is fine classically, we would expect quantum corrections to   introduce some Stuckelberg charge as there is no symmetry to prevent it. The typical value of $\Gamma$ ought to be non-zero, but suppressed by some power of the Lorentz symmetry breaking scale $M_{UV}$.
\section{The Ugly (or the Beautiful?)} \label{sec:ug}
Up until now our discussion has focussed on the non-projectable version of Ho\v rava gravity, in its original manifestation \cite{Horava1}.  In an attempt to alleviate some of the more serious problems  I have discussed (such as dynamical inconsistency and strong coupling), Blas {\it et al} proposed an extension of this model, including terms in the Lagrangian of the form $a_i=\nabla_i \log N$ \cite{Blas2}.  I refer the reader to \cite{klus} for a discussion on the constraint algebra and dynamical consistency, and focus instead  on explaining the proposed resolution of the strong coupling problem. Recall that the main issue with this was the fact that the Stuckelberg action looks nothing like the renormalisable actions studied in \cite{Visser1}, and that crucially, the dispersion relation for the Stuckelberg field does {\it not} scale like $z=3$ in the UV \cite{Blas3, me2}. By adding terms to the action like 
\be
\frac{\alpha}{\kappa}\sqrt{g} N  a_i a^i, \qquad  A \sqrt{g} N a_i a_j R^{ij}, \qquad B \kappa a_ia^i a_j a_k R^{jk}
\ee
the  dispersion relation is modified so that it has the correct UV scaling ($w^2 \propto p^6$). Specifically,
\be
w^2=c_s^2 p^2+\frac{p^4}{M_A^2} +\frac{p^6}{M_B^4},
\ee
 where $c_s^2=\frac{\lambda-1}{\alpha}, ~ M_A \sim  \left(\frac{\alpha}{A}\right)^{1/2}M_{pl}, ~M_B \sim \left(\frac{\alpha}{B}\right)^{1/4}M_{pl}$. For $M_B \sim M_{pl}$ strong coupling is still a problem, kicking in at a scale $\Lambda_{strong} \sim (\lambda-1)^{3/4}\alpha^{-1/4}M_{pl}c^2$ \cite{pap, me2}. This comes as no surprise -- to avoid strong coupling problems we need to introduce some new physics below the would be strong coupling scale.  In fact,  we need  to have $M_B \ll \sqrt{\alpha}M_{pl}$ in order that the $z=3$ scaling kicks in before strong coupling \cite{Blas3}. In some ways this proposal is reminiscent of string theory in which the string scale is introduced below the Planck scale where strong coupling would otherwise occur.
 
 The reason we have labelled this version of theory as ugly is that the numbers involved are quite large. To see this note that the condition $M_B \ll \sqrt{\alpha}M_{pl}$ corresponds to $B \gg \alpha^{-1}$. As tests of Lorentz violation require $\alpha \lesssim 10^{-7}$,  we are forced to take $B \gg 10^7$ \cite{Blas3}. Of course, as we stated in the introduction, this would be  a small price to pay for a consistent quantum theory of gravity! We should also point out that it really is just a question of aesthetics -- the choice of $M_B$  parametrically below $M_{pl}$ is not technically unnatural. The point is that  $M_B$ sets the cut off and so neither $M_B$ nor $M_{pl}$ receive large corrections.
 
 So, do we have a consistent quantum theory of gravity? Not necessarily -- it is too early to say. Whilst Blas {\it et al}'s model might cure strong coupling issues, it could do so in such a way that it runs into real trouble with tests of Equivalence Principle. Recall that we expect the typical value of the Stuckelberg charge to be non-zero, but suppressed by some power of the Lorentz symmetry breaking scale $M_{UV}$. The Stuckelberg charges must therefore be relatively large since we need a low scale of Lorentz violation  $M_{UV} \sim M_B \ll M_{pl}$ to avoid strong coupling. As $\lambda$ cannot be made arbitrarily large, it is clear that Blas {\it et al}'s model could face a challenge from EP tests as described at the end of the previous section. It all boils down to the tension in choosing the value of $M_B$: choose too high a value and you get strong coupling, too low a value and you fail EP tests. We are currently working out the details to see if an acceptable choice for $M_B$ really exists \cite{me3}.

 Finally, I would like to highlight a very recent result in which it was claimed that one must take $\lambda < 1/3$ in order to have a stable vacuum in the quantum version of the theory \cite{shu}. It is difficult to see how this can be compatible with the phenomenological requirement that $|\lambda -1| < 10^{-7}$ at low energies. We should not suppose that $\lambda$ runs from $\lambda<1/3$ in the UV to $\lambda=1$ in the IR, since $\lambda=1/3$ is expected to be a fixed point \cite{Horava1}, and the Stuckelberg field is a ghost whenever $1/3< \lambda<1$ \cite{Blas2}.

\ack{I would like to thank my collaborators, both past and present, including Paul Saffin, Gustavo Niz, Christos Charmousis, Thomas Sotiriou and especially Ian Kimpton. I would also like to thank Antonios Papazoglou and Oriol Pujolas for useful discussions. I am funded by a Royal Society University Research Fellowship.}

\end{document}